\documentclass[twocolumn,showpacs,prb,amsmath,amssymb]{revtex4}

\usepackage{graphicx}% Include figure files
\usepackage{amsmath}
\usepackage{amssymb}
\usepackage{bm}
\usepackage{bbm}
\usepackage{mathbbol}

\DeclareMathAlphabet{\mathpzc}{OT1}{pzc}{m}{it}
\DeclareMathOperator{\Real}{Re}

\newcommand{\identity}{\mathbbm{1}}
\newcommand{\bigO}[1]{\ensuremath{{\cal O}\left( #1 \right)}}
\newcommand{\clusterCountW}[1]{\hat{W}^{[#1]}}
\newcommand{\clusterCountWl}[1]{\hat{W}^{[#1]}_l}
\newcommand{\clusterContrib}[1]{\langle \hat{W}_{#1}' \rangle}
\newcommand{\setSoln}[1]{\langle \hat{W}_{#1} \rangle}

\begin{document}

\title{Concatenated dynamical decoupling in a solid-state spin bath}
\author{W. M. \surname{Witzel}} 
\author{S. \surname{Das Sarma}}
\affiliation{Condensed Matter Theory Center,
  Department of Physics, University of Maryland, College Park, MD
  20742-4111} 
\date{\today}
\begin{abstract}
Concatenated dynamical decoupling (CDD) pulse sequences hold much promise as a
strategy to mitigate decoherence in quantum information processing.
It is important to investigate the actual performance of these
dynamical decoupling strategies in real systems that are promising
qubit candidates.  In this Rapid Communication, 
we compute the echo decay of concatenations
of the Hahn echo sequence for a solid-state electronic 
spin qubit in a nuclear
spin bath using a cluster expansion technique.  
We find that each level of concatenation reverses the effect of 
successive levels of intrabath fluctuations.
On the one hand, this advances CDD as a versatile and realistic
decoupling strategy. 
On the other hand, this invalidates, as overly optimistic, results of
the simple pair approximation used
previously to study 
restoration, through CDD, of coherence lost to a mesoscopic spin bath. 
\end{abstract}

\maketitle

Successful large-scale quantum computation which relies exlusively on
quantum error correcting codes 
necessitates extremely
weak quantum decoherence ($< 10^{-4}$); this is difficult, if not
 impossible, to achieve in the strongly coupled solid-state
 environment.  This stringent requirement arises from the fact that
 the quantum error correction (QEC) overhead grows exponentially (e.g., the
 number of auxiliary qubits needed for encoding) with the magnitude of
 decoherence.
At a local level, decoherence can be mitigated by logically encoding
qubits in a subspace that is less sensitive to noise and by applying dynamical
decoupling (DD) controls.
To meet QEC requirements, it is likely necessary to
combine logical qubit encoding and DD in solid-state 
systems.\cite{ByrdLidarPRL}
We focus our study on DD, which may
employ deterministic,\cite{ViolaPRL,
ViolaPRA, Khodjasteh}
random,\cite{randomDD} or hybrid deterministic-random\cite{hybridDD} 
sequence of qubit controls to partially reverse the
effects of decoherence.
Deterministic sequences hold promise 
for slow, predictable bath systems such as the nuclear spin bath that we consider.
In a universal DD sequence, the time-averaged
Hamiltonian decouples the system from the bath, destroying the 
lowest-order system-bath coupling in a Magnus\cite{Magnus}
expansion.\cite{Khodjasteh}
The Hahn echo, with a single $\pi$ pulse applied midway through the
evolution, is a simple universal DD sequence for a dephasing qubit.
Periodic DD (PDD), repetitive applications of universal DD 
proven effective for nuclear magnetic resonance purposes for many
years,\cite{Meiboom} simply trades an increase in the number of
applied pulses for an increase in their frequency. 
Higher orders of the Magnus expansion, however, limit the performance
of PDD; it is possible to overcome this effect using concatenated DD
(CDD), recursive applications of a universal DD
sequence.\cite{Khodjasteh}

In this Rapid Communication, we
consider an archetypical qubit, a localized electron spin in a
semiconductor at low temperature.
The energy relaxation process of this spin can be made arbitrarily
slow by lowering the
temperature (e.g., $T_1 \gg 1~\mbox{ms}$ for $T \sim 100~\mbox{mK}$) because phonons are
exponentially suppressed below the Bloch-Gruneisen temperature
($\sim 10~\mbox{K}$). 
In the limit of a strong applied magnetic field,
slow fluctuations of the surrounding nuclear spins will induce
electron spin dephasing
(``spectral diffusion''~\cite{deSousaSD}) 
as the primary decoherence mechanism on a time scale of 
$T_2 \sim 1~\mbox{$\mu$s}~\mbox{(GaAs)} -
100~\mbox{$\mu$s}~\mbox{(Si:P)}$.
We show that this decoherence can be corrected, in principle, with
arbitrary precision by using concatenations of the Hahn echo.
The true reason for this, interestingly, goes beyond consideration of orders in the
Magnus expansion; each concatenation of the sequence reverses 
the effect of successive levels of intrabath processes 
(specifically, CDD cancels orders of an intrabath perturbation).

There have been general
theoretical discussions~\cite{Khodjasteh} about using CDD for
qubit preservation without quantitative applications to specific
realistic models of qubits in mesoscopic baths.
To the extent that specific
mesoscopic applications~\cite{yaoCDD} exist for the spin qubit example
considered in this paper, they invariably involve highly simplified and
uncontrolled (as well as untested) models, e.g., 
the nuclear spin pair approximation, which
fails for the realistic spin bath in the
context of CDD.
The pair approximation fails precisely because CDD, beyond canceling
orders of a time (effectively Magnus) expansion,  reverses
the effect of successive levels of intrabath processes.
As these lower levels of interactions are reversed
(e.g., a single flip-flop of a pair of nuclei), 
larger clusters become perturbative equals to pairs and can dominate
decoherence when they are greater in number.
These larger clusters will diminish the effect of CDD
(Refs.~\onlinecite{yaoCDD} give overly optimistic predictions).
Nonetheless, we show that, contrary to PDD, concatenation allows one
to obtain the same fidelity while {\it decreasing} the frequency of
applied pulses.
Our
work is, to the best of our knowledge, the first theoretical
application of the CDD scheme for a realistic situation, establishing,
as a matter of principle, the reasonable prospects for perpetual
coherent control of a qubit.
In proof of concept, we demonstrate this  
in the form of quantum memory retention; however, it is
known~\cite{ViolaPRL} to be possible to effect nontrivial
quantum evolution while performing DD.
In fact, with proper logic encoding such that logical and DD
operations commute, it may be possible~\cite{LidarUpcoming} to perform
CDD and quantum logic simultaneously, 
making CDD a viable strategy to meet the error correction
threshold for a fault-tolerant solid-state spin quantum computer.

%Dynamical decoupling (DD) strategies attempt to preserve a quantum state
%by applying rapid pulses to the quantum system in a way
%that will decouple the system of interest from a decoherence-inducing
%bath.
%Such techniques have developed over many ($\sim$fifty)
%years in the field of nuclear magnetic resonance
%to perform precise spectroscopy on complex molecules
%~\cite{NMR_DD}.
%More recently, DD strategies have been
%discussed~\cite{ViolaPRA, Khodjasteh, otherQC_DD} in the context of
%quantum computing.
%Periodic sequences, known as the
%``bang-bang'' control in quantum information
%literature, prolong overall coherence times 
%by repeating a decoupling pulse sequence faster than the dynamics of the 
%decoherence process~\cite{ViolaPRA}.
%Concatenated sequences employ a recursive structure in
%time that mimics the spatial concatenation of quantum error 
%correction~\cite{Khodjasteh} and can offer better
%results than periodic sequences~\cite{Khodjasteh, yaoCDD, DobrovitskiCDD}.

%In this work, we study decoherence and DD
%of a solid-state spin qubit in a bath of nuclear spins.

The continuous electron spin dephasing
due to its coupling to the nuclear spin bath is a difficult process to
analyze theoretically, in spite of a very long history,\cite{SDhistory}
due to the quantum, non-Markovian nature of nuclear spin flip-flops.  
We previously\cite{witzelHahnShort, witzelHahnLong} 
developed a quantum theory of
nuclear-induced spectral diffusion using a formally exact 
cluster expansion technique and analyzed the Hahn echo 
decay of
solid-state spin qubits, and later\cite{witzelCPMG} adapted the
technique to study the periodic 
Carr-Purcell-Meiboom-Gill\cite{Meiboom} (CPMG) sequence.
Concatenations of the Hahn echo sequence were analyzed in 
Refs.~\onlinecite{yaoCDD} for mesoscopic quantum-dot baths, 
which, however, employed a pair approximation that is not accurate 
in light of the perturbative cancellations that CDD sequences induce.

%We consider a solid-state spin qubit in a nuclear spin bath.  
Specifically,
we treat a donor electron spin in Si:P and a quantum dot
electron spin in GaAs.  Our electron spin qubit will precess in an
applied magnetic field ($\sim 1~\mbox{T}$) 
at its Zeeman frequency ($\sim 10~\mbox{GHz}$) 
with some fluctuations
due to hyperfine (hf) interactions ($\sim 100~\mbox{kHz}$) with a 
nuclear spin bath.  
%The nuclei
%also have Zeeman interactions ($\sim 10~\mbox{MHz}$) with the magnetic field.  
The strong electron Zeeman energies compared with hf and 
nuclear Zeeman energies ($\sim 10~\mbox{MHz}$) suppresses direct hf interactions that would
flip the electron spin.
Most importantly, the nuclei also interact with each other magnetically 
through their dipoles ($\sim 10~\mbox{Hz}$) which, if unchecked, will 
lead to an irreversible loss of information.

A general dephasing Hamiltonian may be
written in the form
$\hat{\cal H} = \sum_{\pm} \lvert \pm \rangle \hat{\cal H}_{\pm} \langle \pm
\rvert$, where $\hat{\cal H}_{\pm}$ acts only upon the bath's Hilbert space.  
We can split $\hat{\cal H}_{\pm}$ into qubit-dependent and -independent
parts, so that 
$\hat{\cal H}_{\pm} = \pm \hat{\cal H}_{qb} + \epsilon \hat{\cal H}_{bb}$,
where $\hat{\cal H}_{qb}$ contains the qubit-bath interaction terms
and $\hat{\cal H}_{bb}$ contains the
intrabath interaction terms.
The $\epsilon$ is a bookkeeping parameter that will be discussed later
in terms of an intrabath perturbation.
From the Fermi-contact hf interaction, 
$\hat{H}_{qb} = \frac{1}{2} \sum_n A_n \hat{I}_{nz}$,
where 
$A_n$ is proportional to the probability of the electron
being at the $n$th nuclear site.
From the secular dipolar interaction, 
$\hat{\cal H}_{bb} = \sum_{n \ne m}' b_{nm}
I_{n+} I_{m-}  - \sum_{n \ne m} 2 b_{nm} I_{nz} I _{mz}$ where the
first summation is restricted to pairs of like nuclei (so that
Zeeman energy is preserved when they flip-flop) and
$b_{nm}$ are dipolar coupling constants.  
We have included the most relevant interactions in the limit of a
strong applied field as discussed in detail in Ref.~\onlinecite{witzelHahnLong}.

Given an
$\hat{\cal H} = \sum_{\pm} \lvert \pm \rangle \hat{\cal H}_{\pm} \langle \pm
\rvert$, the evolution operator $\hat{U}$ for any
sequence of $\pi$ pulses (acting perpendicularly to the
applied field) may be written in the form 
$\hat{U} = \sum_{\pm} \lvert
\pm \rangle \hat{U}^{\pm} \langle \pm \rvert$ 
(or $\hat{U} = \sum_{\pm} \lvert
\mp \rangle \hat{U}^{\pm} \langle \pm \rvert$) for an even (or odd)
number of applied pulses. 
With the up-down free evolution operators denoted 
$\hat{U}_0^{\pm}(\tau) = \exp{(i \hat{\cal H}_{\pm} \tau )}$, 
the Hahn echo sequence 
of $\tau \rightarrow \pi
\rightarrow \tau \rightarrow \pi$, with an extra $\pi$ pulse at the end to
return the electron spin to its original
orientation, has up-down evolution operators of 
$\hat{U}_{1}^{\pm} = \hat{U}_0^{\mp}(\tau) \hat{U}_0^{\pm}(\tau)$.
This trivially modified Hahn echo 
is the first level of the CDD series we consider.  
The Hahn echo sequence is best known as a 
means~\cite{SDhistory}
to eliminate the effects of inhomogeneous broadening (reversing the
effects of a distribution of electron Zeeman frequencies in an
ensemble).  However, the Hahn echo is also an
elementary DD sequence because the time-averaged Hamiltonian,
proportional to $\hat{\cal H}_{+} + \hat{\cal H}_{-} \propto \hat{\cal
  H}_{bb}$, involves no qubit-bath interactions
(in this respect, it does not yield a proper $T_2$ time consistent with the free
induction decay of a single qubit\cite{yaoHahn}). 
As a DD sequence, it may be concatenated in hopes of
successively improving qubit preservation times.\cite{Khodjasteh}
At the $l$th level, our CDD pulse sequence
for $l>0$ is
$\mbox{p}_l := 
\mbox{p}_{l-1} \rightarrow \pi \rightarrow \mbox{p}_{l-1} \rightarrow \pi$,
and $\mbox{p}_0 := \tau$~\cite{Khodjasteh}.
With each concatenation, we do to the previous sequence what the Hahn
echo does to free evolution and in this way obtain improved DD.
This sequence may be simplified by noting that two $\pi$ pulses in
sequence do nothing.  Thus,
\begin{equation}
\label{concatenatedPulseSequence}
\mbox{p}_l := \left\{ 
\begin{array}{ll}
\mbox{p}_{l-1} \rightarrow \pi \rightarrow \mbox{p}_{l-1} &,~ \mbox{odd}~l \\
\mbox{p}_{l-1} \rightarrow \mbox{p}_{l-1} &,~ \mbox{even}~l 
\end{array}\right.,
\end{equation}
and the up-down evolution operators at level $l$ have the recursive
form of~\cite{yaoCDD}
\begin{equation}
\label{Ul_recursive}
\hat{U}_l^{\pm} = \hat{U}_{l-1}^{\mp} \hat{U}_{l-1}^{\pm}.
\end{equation}

In order to characterize the coherence decay, we consider
the transverse component of the
qubit's expectation value; 
normalized to a maximum of $1$, the pulse sequence echo $v_E$ is
defined in this way such that
$v_E = \| \langle 
[\hat{U}^-]^{\dag} \hat{U}^+ \rangle \|
=  \| \langle \hat{W} \rangle \|$
where $\hat{W} \equiv [\hat{U}^-]^{\dag} \hat{U}^+$ and
the $\langle ... \rangle$ denotes an appropriately weighted
average over the bath states.
We assume that the bath is
fully random and use equal weights in averaging over bath states.\cite{footnote_randomBath}

We can solve this mesoscopic quantum problem using a cluster
method.\cite{witzelHahnShort, witzelHahnLong}  
Consider expanding $\hat{W}$ such that
 $\hat{W} = \sum_{n=0}^{N} \clusterCountW{n}$, where $\clusterCountW{n}$ contains
contributions to $\hat{W}$ that involve $n$ separate clusters of
``operatively'' interacting nuclei.  
To be specific, the set of nuclei involved in a term of
$\clusterCountW{1}$ must all be connected together via factors of bilinear
interaction operators to form a single connected cluster.
Clusters have spatial proximity when interactions are local.
If it is possible to approximate $\langle \clusterCountW{1} \rangle$
by including only clusters up to some small size that is much less
than the number of nuclei in the bath, $N$, and if the bath is
effectively uncorrelated initially (e.g., a random bath), then
$\langle \clusterCountW{n} \rangle \approx \langle \clusterCountW{1}
\rangle^n / n!$.
In this ``cluster approximation,''
\begin{equation}
\label{clusterApprox}
v_E = \| \langle \hat{W} \rangle \| \approx 
\exp{(\Real\{ \langle \clusterCountW{1} \rangle \})}.
\end{equation}
This approximation can be tested 
by quantifying the contributions of extraneous terms due to overlapping
clusters when distributing through $\langle \clusterCountW{1}
\rangle^n$.\cite{witzelHahnLong} 
Formally, $\langle \clusterCountW{1} \rangle$ is the sum of all
cluster contributions, $\langle \clusterCountW{1} \rangle = 
\sum_{\cal C} \clusterContrib{\cal C}$, where the contribution
from some cluster ${\cal C}$ may be computed using the recursive formula
\begin{equation}
\label{vC'_recursive}
\clusterContrib{\cal C} = \setSoln{\cal C} -
\sum_{
\substack{
\left\{{\cal C}_i\right\}~\mbox{\scriptsize{disjoint}}, \\
{\cal C}_i \ne \emptyset,~{\cal C}_i~\subset~{\cal C}
}}
\prod_i \clusterContrib{{\cal C}_i},
\end{equation} 
and $\setSoln{\cal C}$ is simply the solution to $\langle \hat{W}
\rangle$ including only the nuclei in the set ${\cal C}$.
The cluster expansion results from expanding $\langle
\clusterCountW{1} \rangle$ 
in Eq.~(\ref{clusterApprox}) to include clusters of successively increasing
size. 

\begin{figure}
\begin{center}
\includegraphics[width=3in]{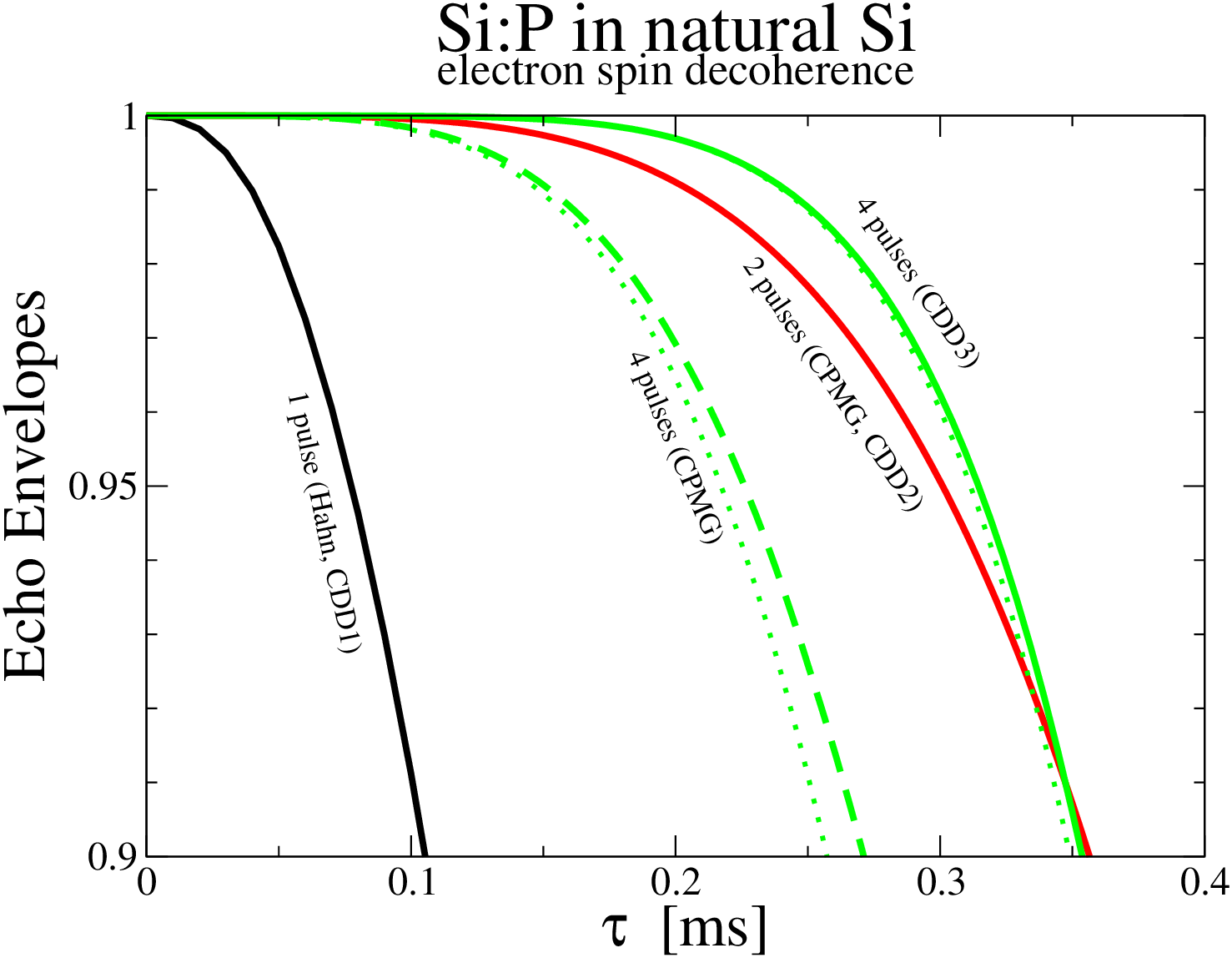}
\end{center}
\caption[Theoretical periodic and concatenated pulse sequence echoes of a donor
  electron in Si:P.]{
Echo decay of an electron bound to a P donor in natural Si 
with an applied magnetic field along the [100] lattice direction
for
different DD pulse sequences as a function
of the minimum time $\tau$ between consecutive pulses.
The dotted lines give corresponding results to the lowest order in
the intrabath perturbation.  
Only the first $90\%$ of the decay is shown to
avoid displaying regions in which the cluster expansion fails to converge.  
Conveniently, coherence much better than $90\%$ is
necessary for QEC protocols and the long-time
coherence decay is irrelevant.
\label{SiPCDDvsCPMG}}
\end{figure}

\begin{figure}
\begin{center}
\includegraphics[width=3in]{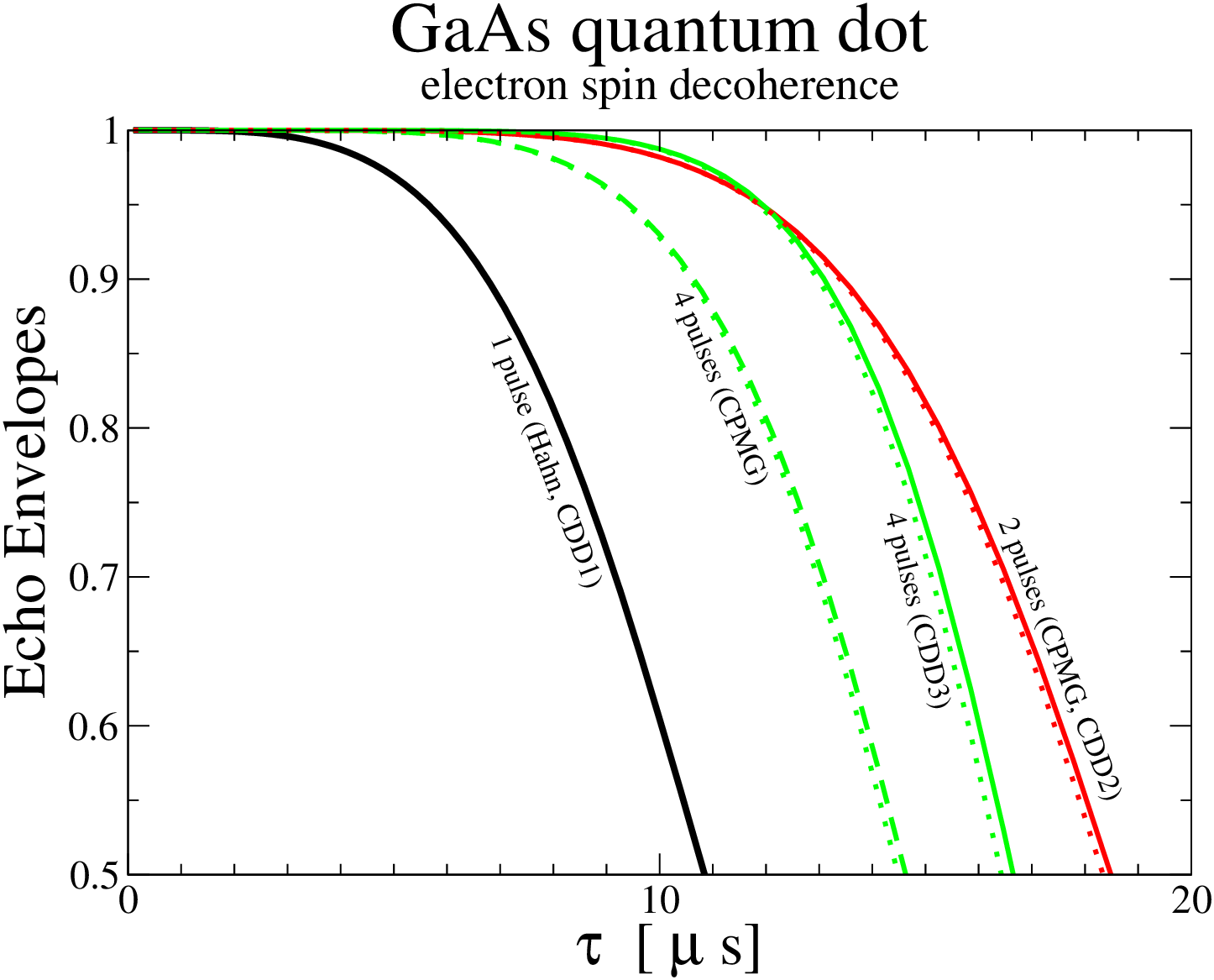}
\end{center}
\caption[Theoretical periodic and concatenated pulse sequence echoes in GaAs
  quantum dots.]{
Echo decay of a quantum dot 
(Fock-Darwin radius of $\ell = 25~\mbox{nm}$ and quantum well thickness
of $z_0 = 8.5~\mbox{nm}$)
electron in GaAs
with an applied magnetic field along the [110] lattice direction
for different DD pulse sequences as a function
of the minimum time $\tau$ between consecutive pulses.  
The dotted lines give corresponding results to the lowest order in
the intrabath perturbation.  
\label{GaAsCDDvsCPMG}}
\end{figure}

We make a comparison between our convergent cluster expansion results
for CDD and earlier calculated\cite{witzelCPMG} CPMG pulse 
sequence echoes of both an electron bound to a P donor in natural
Si (Fig.~\ref{SiPCDDvsCPMG}) and a quantum dot electron in GaAs
(Fig.~\ref{GaAsCDDvsCPMG}) 
plotted as a function of the minimum time $\tau$ between
consecutive pulses.  
The two-pulse CPMG sequence is the same as the CDD2 sequence ($l=2$
CDD), and longer even-pulsed CPMG sequences are simple repetitions of
the CDD2 sequence.
The notable improvement of even-pulsed CPMG sequences over the Hahn
echo\cite{witzelCPMG} is directly due to its extra level of concatenation.
With an increasing number of CPMG pulses, the coherence as a function
of $\tau$ is diminished, but, as noted in Ref.~\onlinecite{witzelCPMG}, the
coherence as a function of the total sequence time, 
$t = 4 \nu \tau$ for a sequence with $2 \nu$ pulses, tends to improve.  
With each level of
concatenation in the CDD series, 
short-time coherence as a function of either $\tau$ or the total sequence time
$t = 2^{l+1} \tau$ improves.  This demonstrates the restorative power of concatenation.

We can understand the reason for 
CDD coherence restoration 
as successive cancellations in perturbative orders.
In a perturbation whose order increases with increasing cluster size,
the lowest order of $\langle \hat{W} \rangle - 1$ is equivalent to the
same order of $\langle \clusterCountW{1} \rangle$ 
because terms of $\hat{W}$ with
multiple clusters are automatically higher-order terms (being products
of lower-order terms).  
This is formal equivalence that we use for convenience only (the
entire purpose of the cluster expansion is to avoid applying such a
perturbative expansion to $\langle \hat{W} \rangle$ directly because it would
fail to converge due to the large size of the bath). 

Using the unitarity property of the evolution operators
in $\hat{W}_l =
[\hat{U}^{-}_l]^{\dag} \hat{U}_l^{+}$ 
and defining $\Delta_l \equiv
\hat{U}^{+}_l - \hat{U}^{-}_l$, 
\begin{eqnarray}
\Real{\left\{\langle \hat{W}_l \rangle\right\}}
% &=& 
%\frac{1}{2} \left\langle \left[\hat{U}^{-}_l\right]^{\dag} \hat{U}_l^{+}
%+ \left[\hat{U}_l^{+}\right]^{\dag} \hat{U}_l^{-} \right\rangle  \\
&=& 1 - \frac{1}{2} \langle\Delta_l^{\dag} \Delta_l \rangle, 
\end{eqnarray}
so that $\Real{\{ \langle \hat{W}_{l} \rangle \}} - 1 =
-\bigO{\Delta_l^2} = \Real{\{ \langle \clusterCountWl{1} \rangle
  \}}$ and $v_E = 1 - \bigO{\Delta_l^2}$.
Applying the recursive definitions for the $\hat{U}^{\pm}_l$ evolution
operators [Eq.~(\ref{Ul_recursive})],
\begin{equation}
\label{Delta_l_Def}
\hat{\Delta}_l \equiv \hat{U}_l^{+} - \hat{U}_l^{-} = [\hat{U}_{l-1}^{-},
  \hat{U}_{l-1}^{+} ] = [\hat{U}_{l-1}^{-}, \hat{\Delta}_{l-1}],
\end{equation}
noting that $\hat{U}^{-}_{l-1}$ commutes with itself.

Consider a perturbation with a smallness parameter $\lambda$ in
which $\hat{U}_l^{\pm} = \identity + \bigO{\lambda}$ for all $l \geq
l_0$ for some $l_0$.
Because the identity commutes with anything,
Eq.~(\ref{Delta_l_Def}) implies that
$\hat{\Delta}_l = \bigO{\lambda} \times \hat{\Delta}_{l-1}$ for all $l >
l_0$.
A perturbation in the time parameter $\tau$ as well as a perturbation in the
intrabath interactions $\epsilon$ both have this property 
with $l_0 = 0$ and $l_0 = 1$, respectively; 
CDD therefore cancels successive orders in $\tau$ and
$\epsilon$. 
The successive cancellation in the time perturbation was previously
noted in Refs.~\onlinecite{yaoCDD}, but not that in the intrabath perturbation.
Both of these perturbations are in correspondence with the
cluster expansion because an increase in cluster size necessitates an
increase in the minimum number of bilinear interaction factors 
$\epsilon$, and a corresponding increase in $\tau$ factors (by
time-energy dimensional arguments). 
In fact, the cluster expansion converges due to the $\epsilon$
perturbation (the time perturbation is additionally
applicable only in special cases) 
as demonstrated in Figs.~\ref{SiPCDDvsCPMG} and
\ref{GaAsCDDvsCPMG} where the dotted lines give the results with
$\langle \clusterCountW{1} \rangle$ approximated by the lowest order in $\epsilon$.
This suggests that CDD provides relevant
perturbative cancellations in the same regime where the cluster expansion
converges.

\begin{figure}
\begin{center}
\includegraphics[width=3in]{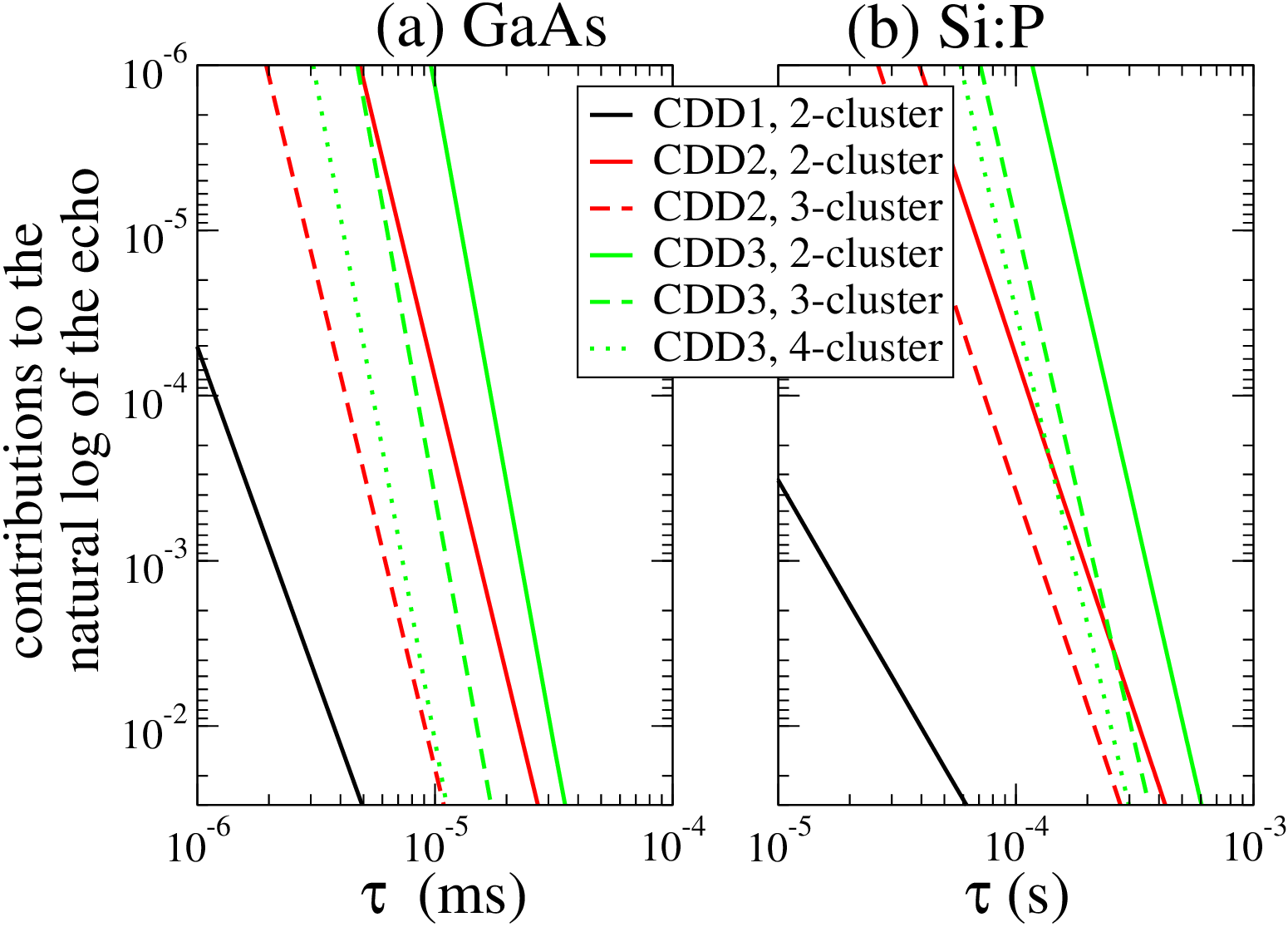}
\end{center}
\caption
{Contributions from clusters of different sizes for the concatenated echo
decays in the (a) Si:P system of Fig.~\ref{SiPCDDvsCPMG} and
(b) GaAs system of Fig.~\ref{GaAsCDDvsCPMG}.
The minimum cluster size required to yield the appropriate
lowest-order result in the cluster expansion increases with each
concatenation, and the larger clusters tend to dominate the decoherence.
\label{CDD_Clusters}}
\end{figure}

With each perturbative cancellation, clusters of increasing size must
be included in the lowest-order cluster expansion calculation as all
have the same perturbative order.
Contributions from different cluster sizes are shown for the CDD
series in Fig.~\ref{CDD_Clusters} for
the Si:P and GaAs systems.  
The larger clusters of the same perturbative order are seen to dominate the
decoherence because they are greater in number
(concatenation of level $l$ is dominated by clusters of size $l+1$), 
invalidating the pair approximation used in Refs.~\onlinecite{yaoCDD}.
This is unfortunate since the pair approximation is simple and
physically transparent.

In summary, we establish the effectiveness of CDD in correcting
spectral-diffusion-induced quantum decoherence in the central spin
problem.  For our problem (and assuming ideal pulses), CDD effectively
restores phase coherence when the interpulse delay is short compared
with the typical intrabath interaction time scale.  Furthermore, each
level of concatenation involves larger clusters of bath spins, thus
invalidating the simple pair approximation used earlier in the
literature.  In reality, the deviation from pulse ideality will
determine the level of feasible quantum error correction; there is,
however, promising recent work\cite{UhrigOptPulse} to address this issue.

Our study of spin bath decoherence provides important
tests for CDD strategies and takes these strategies
beyond the level of abstract formalism\cite{ViolaPRA, Khodjasteh} and 
small-scale models,\cite{Khodjasteh, DobrovitskiCDD}
establishing that CDD could play a key role in correcting
decoherence-induced errors in solid-state spin quantum computer architectures.
An ability to
prolong quantum coherence indefinitely using CDD techniques could lead
to an
effective solid-state spin quantum memory as well as much reduced
overhead
in quantum error correction protocols.

We would like to acknowledge helpful comments from David Lidar and
Kaveh Khodjasteh.
This work is supported by DTO-ARO and NSA-LPS.


\begin{thebibliography}{99}
%
\bibitem{ByrdLidarPRL}
M. S. Byrd and D. A. Lidar, Phys. Rev. Lett. {\bf 89}, 047901 (2002);
Lian-Ao Wu, M.S. Byrd, and D.A. Lidar, {\it ibid.} {\bf 89}, 127901 (2002). 
%
\bibitem{ViolaPRL}
L. Viola, E. Knill, and S. Lloyd, Phys. Rev. Lett. {\bf 82}, 2417 (1999);
%
\bibitem{ViolaPRA}
L. Viola and S. Lloyd, Phys. Rev. A {\bf 58}, 2733
  (1998).
\bibitem{Khodjasteh}
K. Khodjasteh and D.A. Lidar, Phys. Rev. Lett. {\bf 95}, 180501
(2005); Phys. Rev. A {\bf 75}, 062310 (2007).
\bibitem{randomDD}
L. Viola and E. Knill, Phys. Rev. Lett. {\bf 94}, 060502 (2005). 
%
\bibitem{hybridDD}
O. Kern and G. Alber, Phys. Rev. Lett. {\bf 95}, 250501 (2005).
%
\bibitem{Magnus} W. Magnus, Commun. Pure Appl. Math. {\bf 7}, 649
  (1954).
%
\bibitem{Meiboom} H.Y. Carr and E.M. Purcell, Phys. Rev. {\bf 94},
  630 (1954); S. Meiboom and D. Gill, Rev. Sci. Instrum. {\bf 29},
  6881 (1958).
%
%
\bibitem{deSousaSD} R. de Sousa and S. Das Sarma, Phys. Rev. B {\bf 68},
115322 (2003).
%
\bibitem{yaoCDD}
Wang Yao, Ren-Bao Liu, and L. J. Sham,
Phys. Rev. Lett. {\bf 98}, 077602 (2007);
Ren-Bao Liu, Wang Yao, and L. J. Sham, New J. Phys. {\bf 9}, 226 (2007).
%
\bibitem{LidarUpcoming}
D. Lidar (private communication).
%
%
\bibitem{SDhistory}
 B. Herzog and E.L. Hahn, Phys. Rev. {\bf 103}, 148
(1956); A.M. Portis, {\it ibid.} {\bf 104}, 584 (1956);
J.R. Klauder and P.W. Anderson, {\it ibid.} {\bf
  125}, 912 (1962);
C.P. Slichter, {\it Principles of Magnetic
    Resonance}, 3rd ed. (Springer-Verlag, Berlin, 1990).
%
%
\bibitem{witzelHahnShort} W.M. Witzel, R. de Sousa, and S. Das Sarma,
Phys. Rev. B {\bf 72}, 161306(R) (2005).
%
\bibitem{witzelHahnLong}
W.M. Witzel and S. Das Sarma,
Phys. Rev. B {\bf 74}, 035322 (2006).
%
\bibitem{witzelCPMG}W.M. Witzel and S. Das Sarma, 
Phys. Rev. Lett. {\bf 98}, 077601 (2007).
%
\bibitem{yaoHahn}Wang Yao, Ren-Bao Liu, and L. J. Sham,
Phys. Rev. B {\bf 74}, 195301 (2006).
%
\bibitem{footnote_randomBath}
In terms of temperature, dipolar interactions have $\mbox{nK}$ scale and
nuclear Zeeman energies, at a field around $1~\mbox{T}$, 
 have $\mbox{mK}$ scale.  
The random bath assumption is therefore justified for typical thermal baths
of $0.1-1~\mbox{K}$.
Below the $\mbox{mK}$ range, one may
simply weight the average according to nuclear polarization.
%
\bibitem{UhrigOptPulse}
S. Pasini, T. Fischer, P. Karbach, and G. S. Uhrig, arXiv:0709.0588.
%
\bibitem{DobrovitskiCDD}Wenxian Zhang, V. V. Dobrovitski, Lea
  F. Santos, Lorenza Viola, and B. N. Harmon, Phys. Rev. B {\bf 75}, 201302(R)
  (2007).
%
%\bibitem{footnote1}
%Other intra-nuclear interactions, such as indirect
%exchange~\cite{indirectEx}, may also
%play a role and can have the same order of magnitude strength as
%dipolar interactions in GaAs; although any such interaction may be easily
%incorporated into our framework, we have neglected these interactions
%as they have only a quantitive affect that is relatively minor and is
%not consequential to main discussion of this Letter.  
%
\end{thebibliography}
\end{document}